\documentclass{PoS}

\usepackage{amsmath}

\title{Transverse momentum of partons:\\ from low to high pT}

\ShortTitle{Transverse momentum of partons: from low to high pT}

\author{\speaker{Markus Diehl}\\
        Deutsches Elektronen-Synchroton DESY, 22603 Hamburg, Germany\\
        E-mail: \email{markus.diehl@desy.de}}

\abstract{%
  Transverse-momentum spectra in hard processes are typically described
  either in terms of intrinsic transverse momentum of partons, or in terms
  of perturbative radiation.  The relation between these descriptions is
  discussed for the example of semi-inclusive deep inelastic scattering,
  with special focus on the angular distribution of the observed hadron.
  This involves nontrivial theoretical issues, such as the proper
  definition of transverse-momentum dependent parton distributions, and
  has practical consequences for the description of $p_T$ spectra in
  phenomenology.}

\FullConference{LIGHT CONE 2008 Relativistic Nuclear and Particle Physics\\
                 July 7-11, 2008\\
                 Mulhouse, France}

\newcommand{\ms}{\mskip 1.5mu}

\begin{document}

\section{Motivation}

This talk is concerned with the transverse-momentum dependence of
particles produced in scattering processes involving a hard momentum
scale.  Specifically, I will discuss semi-inclusive deep inelastic
scattering, $e p \to e + h + X$, where the momentum of the hadron $h$ is
measured.  Using crossing symmetry, the results can be carried over to
Drell-Yan or $W$ or $Z$ production in $pp$ and $p\bar{p}$ collisions, as
well as to hadron pair production in electron-positron annihilation, $e^+
e^- \to h_1 + h_2 + X$.  Details can be found in the recent paper
\cite{Bacchetta:2008xw}.

\begin{figure}[h]
\begin{center}
\includegraphics[width=0.43\textwidth]{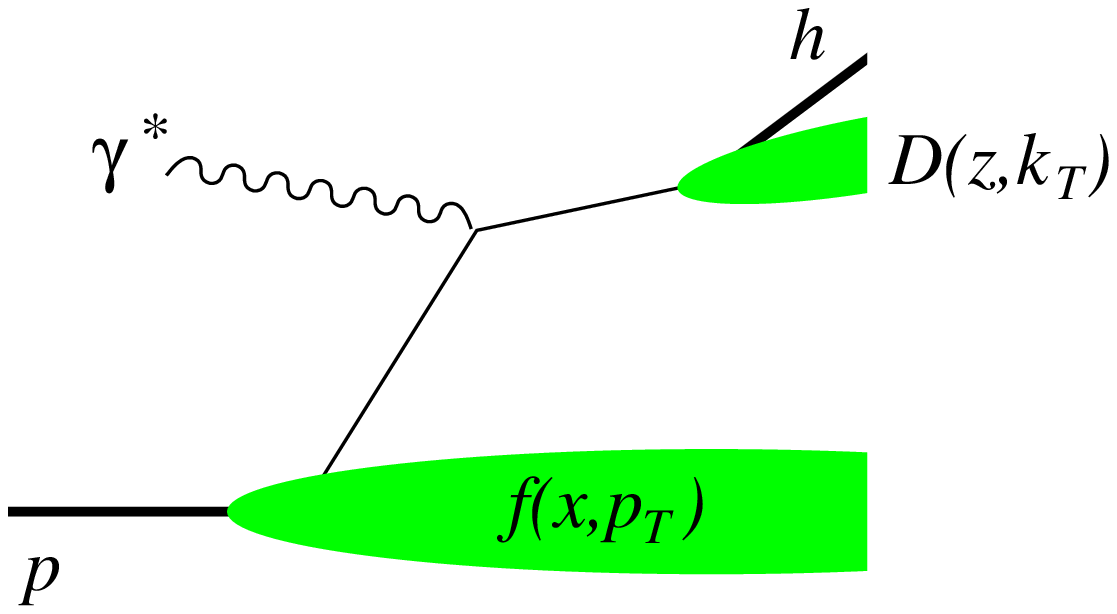}
\hspace{2.1em}
\includegraphics[width=0.43\textwidth]{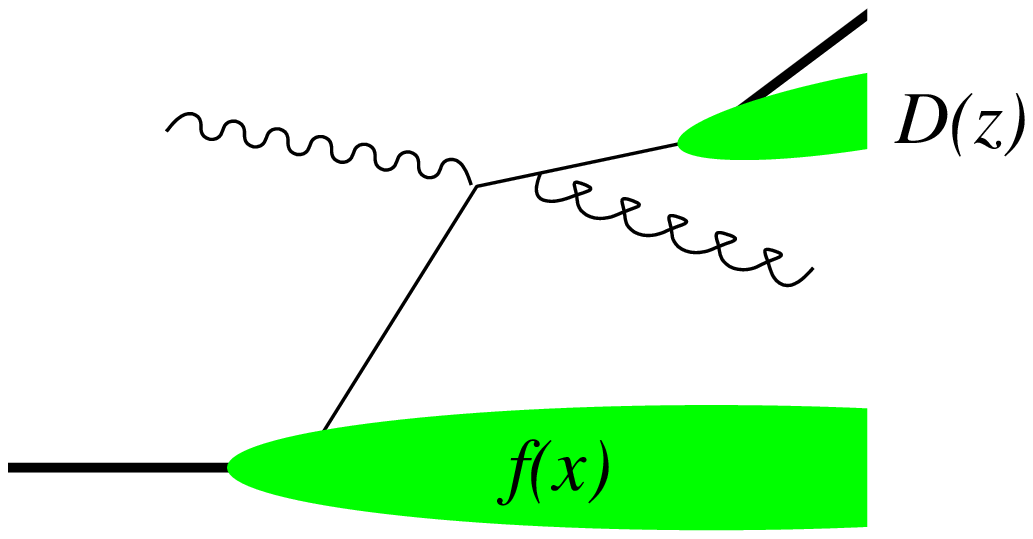} \\[-0.3em]
$\mathbf{(a)}$ \hspace{0.47\textwidth} $\mathbf{(b)}$
\end{center}
\caption{\label{fig:mechanisms} The low-$q_T$ description $\mathbf{(a)}$
  and the high-$q_T$ description $\mathbf{(b)}$ for the
  transverse-momentum distribution of the produced particle $h$ in
  semi-inclusive deep inelastic scattering, $e p \to e + h + X$.}
\end{figure}

The transverse-momentum spectrum of produced particles may be regarded as
a basic feature of the final state.  Even in the simple processes just
mentioned, its investigation reveals a number of nontrivial aspects of QCD
dynamics.  There are two theoretical frameworks to describe the
distribution of a suitably defined transverse momentum $\vec{q}_T$ in the
final state.
The description sketched in Fig.~\ref{fig:mechanisms}a is based on the
``intrinsic transverse momentum'' of partons within a hadron and uses
transverse-momentum dependent (i.e.\ unintegrated) parton densities and
fragmentation functions.  This description can be used for $q_T \ll Q$,
where $Q$ is the virtuality of the photon or electroweak boson, which I
assume to be large throughout this talk.
The description represented in Fig.\ref{fig:mechanisms}b uses collinear
(i.e.\ $k_T$ integrated) parton densities and fragmentation functions,
generating finite $q_T$ by perturbative radiation of partons into the
final state.  This description is adequate for $q_T \gg M$, where $M$
stands for a generic nonperturbative scale.
In the following I refer to the two descriptions as ``low-$q_T$'' and
``high-$q_T$'', respectively.  It has long been known that both mechanisms
give rise to a nonzero cross section for longitudinal photon polarization
and to a nontrivial dependence on the azimuth of $\vec{q}_T$
\cite{Cahn:1978se,Georgi:1977tv,Oganesian:1997jq}.

\section{Insight from power counting}
\label{sec:power}

It is natural to ask how these two descriptions are related to each other.
A first answer can be obtained from a careful look at the power counting
in the region of intermediate transverse momentum, $M \ll q_T \ll Q$,
where both approaches can be applied.  The low-$q_T$ approach starts with
an expansion in the small parameter $q_T/Q$ and involves coefficients
depending on $M/q_T$, which for $q_T \gg M$ can be further expanded in
$M/q_T$.  For an observable $F$ with mass dimension~$-2$ one thus has
\begin{align}
  \label{pow-low}
F(q_T,Q)  \;\stackrel{q_T \ll Q}{=}\; \frac{1}{M^2}\,
  \sum_{\text{twist}~n}\, 
  \biggl[\frac{q_T}{Q}\biggr]^{n-2}\,
  l_{n} \biggl(\frac{M}{q_T}\biggr)
& \;\stackrel{M \ll q_T \ll Q}{=}\; \frac{1}{M^2}\,
  \sum_{n,k}  l_{n,k}\;
  \biggl[\frac{q_T}{Q}\biggr]^{n-2}\,
  \biggl[\frac{M}{q_T}\biggr]^{k} \,.
\end{align}
By contrast, the high-$q_T$ approach first expands an observable in
$M/q_T$, with coefficients that for intermediate $q_T$ can be further
expanded in $q_T/Q$\,:
\begin{align}
  \label{pow-hi}
F(q_T,Q)  \;\stackrel{M \ll q_T}{=}\;  \frac{1}{M^2}\,
  \sum_{\text{twist}~n}\,
  \biggl[\frac{M}{q_T}\biggr]^{n}\,
  h_{n} \biggl(\frac{q_T}{Q}\biggr)
& \;\stackrel{M \ll q_T \ll Q}{=}\;  \frac{1}{M^2}\,
  \sum_{n,k} h_{n,k}\;
  \biggl[\frac{M}{q_T}\biggr]^{n}\,
  \biggl[\frac{q_T}{Q}\biggr]^{k-2} \,.
\end{align}
The simultaneous validity of both approaches in the region $M \ll q_T \ll
Q$ implies $l_{n,k} = h_{k,n}$.  The first index $n \ge 2$ in each
expansion characterizes the twist of the corresponding calculation.  In
practice only terms with $n=2$ and possibly $n=3$ can actually be
calculated.

For observables with nonzero $l_{2,2} = h_{2,2}$, the leading terms in the
two calculations coincide for intermediate $q_T$, where they provide
complementary descriptions of the same physics.  One may then try to
construct a smooth interpolation between the two descriptions that is
valid at all $q_T$.
There are, however, observables with $l_{2,2} = h_{2,2} = 0$, where the
leading term $l_{2,4}$ of the low-$q_T$ result is distinct from the
leading term $h_{2,4}$ of the high-$q_T$ result.  With both calculations
only performed at leading-twist accuracy, one can then add their results
at intermediate $q_T$ without double counting; which of them is more
important at given $q_T$ depends on the relative size of $q_T/Q$ and
$M/q_T$.  We will encounter examples for both cases in
section~\ref{sec:compare}.


\section{Structure functions for semi-inclusive deep inelastic scattering}

To describe the kinematics we use the standard scaling variables $x$ and
$z$, the inelasticity $y$, the photon virtuality $Q$, the scaled
transverse momentum $q_T = P_{h\perp} /z$ of the produced hadron in the
$\gamma^*p$ c.m., and the azimuthal angle $\phi$ between the lepton and
hadron planes in that frame.  Precise definitions are given in
\cite{Bacchetta:2008xw}.  The unpolarized cross section can then be
parameterized in the form
\begin{equation}
  \label{str-fcts}
\frac{d\sigma(ep\to ehX)}{d\phi\, dq_T^2\, dx\, dy\, dz} =
  \text{(kin.~factor)}
  \times \Bigl[
    F_{T} + \varepsilon F_{L}
    + \sqrt{2\varepsilon (1+\varepsilon)} \cos\phi\, F^{\cos\phi}
        + \varepsilon \cos 2\phi\, F^{\cos2\phi} \Bigr] \,,
\end{equation}
where the ratio of longitudinal and transverse photon flux is given by
$\varepsilon = (1-y) /(1-y + y^2/2)$ in the Bjorken limit.  The
semi-inclusive structure functions $F_{\ldots}$ depend on $x$, $z$,
$q_T^2$, $Q^2$, and the subscripts $T$ and $L$ are respectively associated
with transverse and longitudinal photon polarization.


\subsection{High-$q_T$ calculation}

The high-$q_T$ calculation gives the structure functions as convolutions
\begin{align}
F_{\ldots} &= \frac{1}{Q^2\ms z^2}\;
\sum_{i,j = q,\,\bar{q},\,g}\;
  \int_{x}^1 \frac{d\hat{x}}{\hat{x}}\,
  \int_{z}^1 \frac{d\hat{z}}{\hat{z}}\, f_1^i\Bigl(\frac{x}{\hat{x}}\Bigr)\,
  D_1^j\Bigl(\frac{z}{\hat{z}}\Bigr)\,
  K^{ij}_{\ldots} \Bigl(\hat{x},\hat{z}, \frac{q_T}{Q}\Bigr)
\end{align}
in longitudinal momentum fractions, where $f_1^i$ and $D_1^j$ respectively
denote the usual unpolarized collinear distribution and fragmentation
functions, and the $K^{ij}_{\ldots}$ are perturbatively calculable
hard-scatter\-ing kernels.  Expanding these kernels in $q_T/Q$, one
readily obtains the result of the high-$q_T$ mechanism for intermediate
$q_T$.  At order $\alpha_s$, all structure functions in \eqref{str-fcts}
contain a term proportional to $f_1(x)\, D_1(z)\, \log (Q/q_T)$ after this
expansion.  Higher orders provide terms going like $\alpha_{s}^n
\log{}^{2m-1} (Q/q_T)$ with $m\le n$.  To obtain a stable perturbative
result in the region where $\log (Q/q_T)$ is large, these logarithms
should be resummed to all orders.  We will come back to this in
sect.~\ref{sec:low}.


\subsection{Low $q_T$}
\label{sec:low}

In the low-$q_T$ description, factorization is fairly well understood at
twist-two accuracy \cite{Collins:1981uk,Ji:2004wu} and leads to a
representation of the form
\begin{align}
  \label{CS-fact}
F_{\ldots} = \sum_{i=q,\,\bar{q}}\, x e_i^2
 & \int d^2 \vec{p}_T^{}\, d^2 \vec{k}_T^{}\, d^2 \vec{l}_T^{}\,
   \delta(\vec{p}_T^{}-\vec{k}_T^{}+\vec{l}_T^{}+\vec{q}_T^{})\,
   w_{\ldots}(\vec{p}_T, \vec{k}_T)\,
   f^i(x,p_T^2)\, D^i(z,k_T^2)\, U({l}_T^{2})
\end{align}
with known functions $w_{\ldots}\ms$, where for simplicity I have omitted
a hard factor representing $\alpha_s$ corrections from virtual graphs.  In
addition to transverse-momentum dependent distribution and fragmentation
functions $f^i(x,p_T^2)$ and $D^i(z,k_T^2)$, the expression
\eqref{CS-fact} contains a soft factor $U(l_T^2)$ describing soft gluon
exchange between partons moving in the direction of the target and partons
moving in the direction of the observed hadron $h$.  At twist-three
accuracy, soft gluon exchange has not been analyzed, so that we do not
have a full understanding of factorization for structure functions going
like $1/Q$.  However, there are detailed calculations at tree level
\cite{Mulders:1995dh,Boer:2003cm}, which give results very similar in form
to \eqref{CS-fact} without the factor $U(l_T^2)$.

\begin{figure}[b]
\begin{center}
\includegraphics[width=0.43\textwidth]{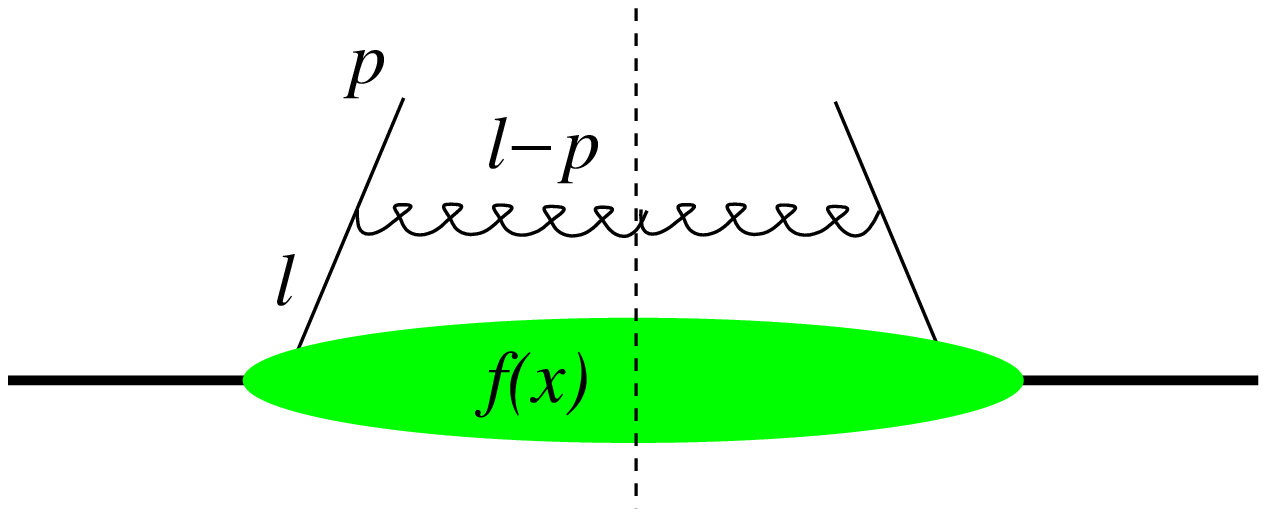}
\hspace{2.1em}
\includegraphics[width=0.43\textwidth]{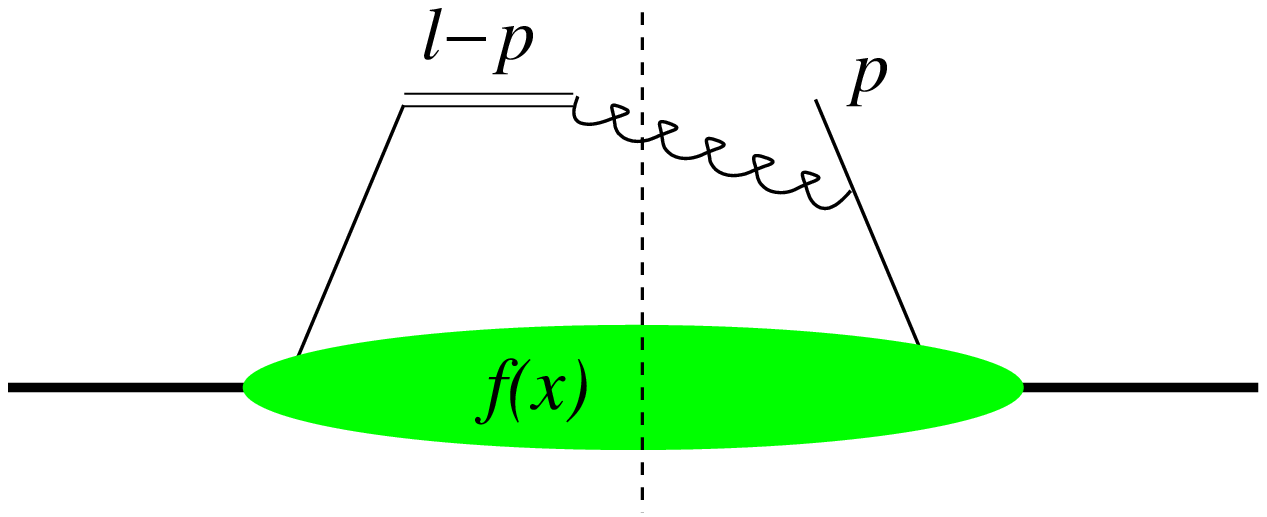} \\[-0.3em]
$\mathbf{(a)}$ \hspace{0.47\textwidth} $\mathbf{(b)}$
\end{center}
\caption{\label{fig:pdfs} Example graphs for the calculation of an
  unintegrated parton distribution at high $p_T$ in terms of a collinear
  distribution $f(x)$ and a hard-scattering subprocess.  The double line
  represents an eikonal propagator, which comes from the Wilson line $P
  \exp\bigl[ -ig \int_0^\infty d\lambda\, n\cdot A(\xi + \lambda n)\bigr]$
  in the definition of the parton density
  \protect\cite{Collins:1981uk,Ji:2004wu,Collins:2008ht}.}
\end{figure}

To evaluate \eqref{CS-fact} for intermediate $q_T$, one notes that for $q_T
\gg M$ at least one of the momenta $p_T$, $k_T$ or $l_T$ has to be large.
In this region, the corresponding factor can be calculated using collinear
factorization, with the large transverse momentum generated by
perturbative parton radiation as illustrated in Fig.~\ref{fig:pdfs}.
As shown in \cite{Bacchetta:2008xw}, general symmetry considerations allow
one to determine the power behavior at high $p_T$ for the different parton
densities parameterizing the spin and momentum dependence of the quark
distribution in a proton.  In particular one finds
\begin{align}
  \label{power-pdfs}
f_1(x,p_T^2)       &\sim \frac{1}{p_T^2}\, \alpha_s\,
          \sum_{i= q,\,\bar{q},\,g}\, [ f_1^i\otimes K_1^i ] \,,
\qquad\qquad
x f^\perp(x,p_T^2)   \sim \frac{1}{p_T^2}\, \alpha_s\,
          \sum_{i= q,\,\bar{q},\,g}\, [ f_1^i\otimes K^{\perp i} ] \,,
\nonumber \\
h_1^\perp(x,p_T^2) &\sim \frac{M^2}{p_T^4}\, \alpha_s\;
          \sum_{i}\,
          [ \text{collinear twist-three distributions} \otimes K_3^i ] \,,
\end{align}
where $K_1^i$, $K^{\perp i}$, and $K_3^i$ are calculable hard-scattering
kernels and $\otimes$ denotes a convolution in longitudinal momentum
fractions.  The Boer-Mulders function $h_1^\perp$ describes the
distribution of transversely polarized quarks in an unpolarized proton,
whereas $f^\perp$ is a distribution of twist three (involving one good and
one bad light-cone component of the quark field).  Explicit calculation of
$f_1(x,p_T^2)$ and $f^\perp(x,p_T^2)$ reveals the rapidity divergences
discussed in \cite{Collins:2008ht}.  They have the form
\begin{equation}
  \label{eikonal}
\frac{1}{(l-p)\cdot n} = \frac{1}{n^-}\,
    \frac{1}{(l-p)^+ + \frac{n^+_{}}{n^-}\, (l-p)^-} 
\qquad\qquad \text{with}~~~
(l-p)^- = \frac{\vec{p}_T^2}{2(l-p)^+}
\end{equation}
and come from the Wilson line in Fig.~\ref{fig:pdfs}b, or equivalently
from the gluon propagator in Fig.~\ref{fig:pdfs}a if one uses the gauge
$n\cdot A=0$ where the Wilson line is unity.  In a parton density for a
fast right-moving proton, the loop variable $l^+$ is integrated down to
its lower kinematic limit $p^+$.  To avoid a logarithmic divergence in
\eqref{eikonal} one must hence keep $n^+$ nonzero (complications arising
for spacelike $n$ are discussed in
\cite{Bacchetta:2008xw,Collins:2008ht}).  The light-cone gauge $A^+=0$ or
a purely lightlike Wilson line cannot be used in this context.  It should
be instructive to investigate how this affects formulations of QCD based
on light-cone gauge.  Note that in the context of light-cone quantization,
configurations with $(l-p)^+ =0$ in Fig.~\ref{fig:pdfs}a correspond to
zero modes of the gluon field.

Keeping $n^+$ finite in \eqref{eikonal} cuts off the region of negative
gluon rapidities, where $(l-p)^+ \to 0$ and $(l-p)^- \to \infty$.  This is
physically reasonable, since fast left-moving gluon modes should not be
included in the parton distribution of a fast right-moving hadron.  The
dependence of unintegrated parton distributions on $n^+/n^-$ is described
by the Collins-Soper equation \cite{Collins:1981uk}, whose solution can be
written as the product of an $n^+/n^-$ independent initial condition and a
Sudakov factor.

Power laws analogous to those in \eqref{power-pdfs} are obtained for the
fragmentation functions $D_1(z,k_T^2)$, $D^\perp(z,k_T^2)$, and the
Collins function $H_1^\perp(z,k_T^2)$.  Together with the perturbative
expression for $U(l_T^2)$ at high $l_T$ one can then determine the
behavior of the semi-inclusive structure functions for $M \ll q_T \ll Q$.
The terms going with $n^+/n^-$ in the distribution and fragmentation
functions give rise to a $\log (Q/q_T)$ in the structure functions, which
we also encountered in the high-$q_T$ calculation.  The Collins-Soper
equation allows one to resum such logarithms to all orders and is at the
origin of the CSS formalism \cite{Collins:1984kg}, which plays a prominent
role in collider phenomenology.  The need to keep $n^+/n^-$ finite in
\eqref{eikonal} is thus not a mere technicality but has practical
implications for physical observables.


\section{Comparing the low- and high-$q_T$ calculations}
\label{sec:compare}

\begin{table}
\begin{center}
\renewcommand{\arraystretch}{1.2}
\begin{tabular}{|r|ccc|cc|} \hline\hline
 & \multicolumn{3}{c|}{low-$q_T$ calculation} &
   \multicolumn{2}{c|}{high-$q_T$ calculation} \\ 
 & power & twist & contributing functions & power & twist \\
\hline
$F_T \sim$ & ${1}/{q_T^2}$ & 2 & {$f_1(x,p_T^2), D_1(z,k_T^2)$}
           & ${1}/{q_T^2}$ & 2 \\
$F_L \sim$ & ${1}/{Q^2}$ & 4 & result unknown
           & ${1}/{Q^2}$ & 2 \\
$F^{\cos 2\phi} \sim$ & ${M^2}/{q_T^4} \phantom{+}$ & 2
 & {$h_1^\perp, H_1^\perp$} & ${1}/{Q^2}$ & 2 \\
 & $+{1}/{Q^2}$ & 4 & result unknown & & \\
$F^{\cos\phi} \sim$ & ${1}/{(Q\ms q_T)}$ & 3 & {$f_1, f^\perp,
   D_1, D^\perp$} & ${1}/{(Q\ms q_T)}$ & 2 \\
\hline\hline
\end{tabular}
\end{center}
\caption{\label{tab:power-results} Behavior of semi-inclusive structure
  functions in the intermediate region $M\ll q_T \ll Q$.}
\end{table}

\begin{figure}[h]
\begin{center}
\includegraphics[width=0.43\textwidth]{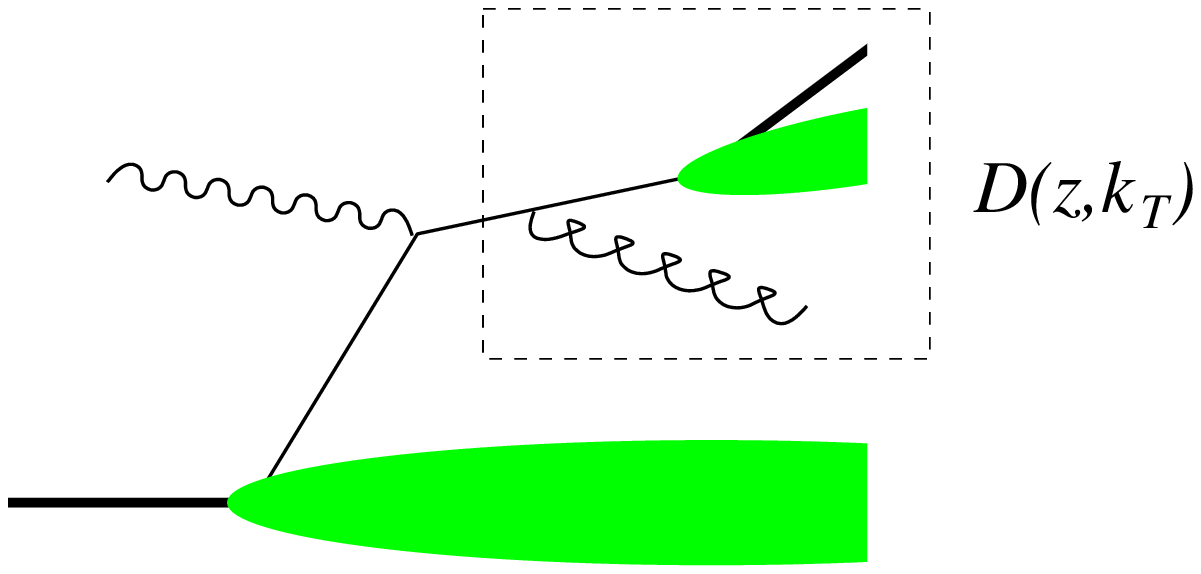}
\hspace{2.1em}
\includegraphics[width=0.40\textwidth]{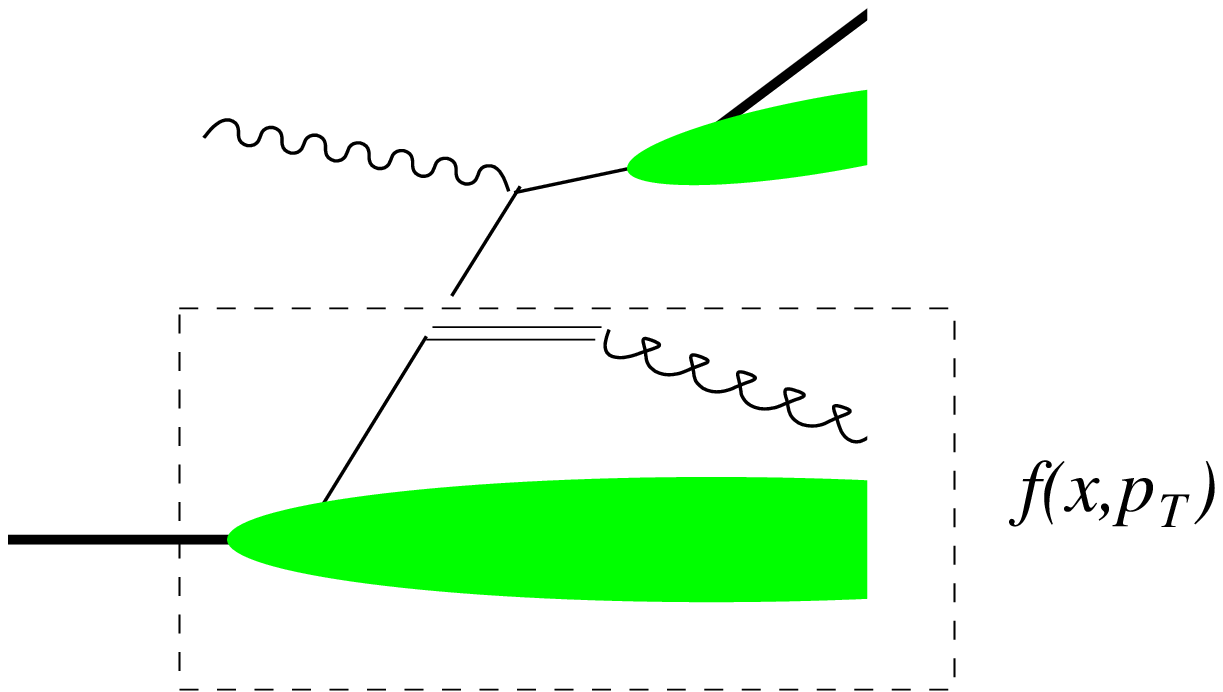} \\[-0.3em]
$\mathbf{(a)}$ \hspace{0.47\textwidth} $\mathbf{(b)}$
\end{center}
\caption{\label{fig:low-qt} Example graphs for the low-$q_T$ calculation
  in the region $q_T\gg M$, where a factorized description as in
  Fig.~\protect\ref{fig:pdfs} is valid for the fragmentation function
  $\smash{\mathbf{(a)}}$ or the parton distribution $\smash{\mathbf{(b)}}$.}
\end{figure}

The behavior for $M\ll q_T \ll Q$ of the unpolarized structure functions
obtained in the low- and high-$q_T$ calculations is given in
Table~\ref{tab:power-results}.  Explicit evaluation shows that for $F_T$
the results of the two calculations exactly coincide; this is an example
of the case where at intermediate $q_T$ the leading term in the expansions
\eqref{pow-low} and \eqref{pow-hi} goes with $l_{2,2} = h_{2,2}$.  The
agreement between the two calculations can be seen at the level of
diagrams: roughly speaking, the graph of Fig.~\ref{fig:mechanisms}b
corresponds to the graph in Fig.~\ref{fig:low-qt}a if the gluon moves fast
in the direction of the hadron $h$, and to the graph in
Fig.~\ref{fig:low-qt}b if the gluon moves fast in the direction of the
target.

The leading term in the low-$q_T$ result for $F^{\cos 2\phi}$ involves the
Boer-Mulders and Collins functions.  This structure function provides an
example for the case where $l_{2,2} = h_{2,2} =0$ and where the leading
contributions $l_{2,4}$ and $h_{4,2}$ in the two calculations are
different and can be added at intermediate $q_T$.
From power counting it is clear that the $1/Q^2$ behavior obtained in the
high-$q_T$ calculation for both $F^{\cos 2\phi}$ and $F_L$ corresponds to
twist-four contributions in the low-$q_T$ framework, whose computation is
well beyond the state of the art.  As a consequence, one cannot invoke the
CSS method \cite{Collins:1984kg} to resum the logarithms in $Q/q_T$ that
appear in the high-$q_T$ result.  Terms going like $1/Q^2$ are obtained if
one calculates $F_L$ and $F^{\cos 2\phi}$ in the parton model
\cite{Cahn:1978se}, considering the graph in Fig.~\ref{fig:mechanisms}a
with only the functions $f_1(x,p_T^2)$ and $D_1(z,k_T^2)$.  However, the
results do not match with the high-$q_T$ calculation at intermediate $q_T$
and can hence only be regarded as partial evaluations of the complete
(unknown) twist-four terms.

If we perform the twist-three calculation for $F^{\cos\phi}$ at low $q_T$
using the tree-level result of \cite{Mulders:1995dh} supplemented with the
soft factor $U(l_T^2)$ of the twist-two factorization formula
\eqref{CS-fact}, we obtain agreement with the high-$q_T$ result for
intermediate $q_T$ \emph{except} for a missing term proportional to
$f_1(x)\, D_1(z)$.  Such a term comes from kinematics where both the plus-
and the minus-momentum of the gluon in Fig.~\ref{fig:low-qt} is
negligible.  This shows that, if a proper factorization formula for twist
three can be established, the soft-gluon sector will have to be treated
with particular care.

\section{Summary}

The descriptions of transverse-momentum spectra based either on intrinsic
transverse momentum of partons or on perturbative radiation are not
disconnected.  They can be related in an intermediate region $M\ll q_T \ll
Q$ by describing transverse-momentum dependent distribution and
fragmentation functions themselves in terms of perturbative radiation, as
indicated in Fig.~\ref{fig:pdfs}.  Understanding the connection between
the two approaches for a given observable enables one to devise
descriptions that may be used in the full region of $q_T$.  As shown in
\cite{Bacchetta:2008xw}, the interplay of the two mechanisms has also
nontrivial consequences in observables that are integrated over $q_T$.

A varied picture arises already for the angular distribution of the
measured hadron in unpolarized semi-inclusive scattering, with the results
obtained at leading-power accuracy in the low- and high-$q_T$ calculations
coinciding for some observables but not for others.  An even richer
phenomenology emerges if one includes polarization effects
\cite{Bacchetta:2008xw,Ji:2006br}.

The calculation of unintegrated parton densities or fragmentation
functions at perturbatively large transverse momenta leads to divergences
from gluonic zero modes if naively performed in light-cone gauge.  Proper
regularization of these divergences physically ensures that left-moving
gluon modes are not included in distribution or fragmentation functions
for right-moving hadrons and provides a powerful method for resumming
logarithms of $Q/q_T$ into a Sudakov factor.  It remains to be understood
how this physics can be treated within light-cone quantization.



\begin{thebibliography}{99}

\bibitem{Bacchetta:2008xw}
  A.~Bacchetta, D.~Boer, M.~Diehl and P.~J.~Mulders,
  \emph{JHEP} {\bf 0808} (2008) 023
  [{\tt arXiv:0803.0227}]. 

\bibitem{Cahn:1978se}
  R.~N.~Cahn,
  \emph{Phys.\ Lett.\  B} {\bf 78} (1978) 269;
%
  \emph{Phys.\ Rev.\  D} {\bf 40} (1989) 3107;\\
%
  M.~Anselmino, M.~Boglione, U.~D'Alesio, A.~Kotzinian, F.~Murgia and
  A.~Prokudin, 
  \emph{Phys.\ Rev.\  D} {\bf 71} (2005) 074006
  [{\tt hep-ph/0501196}].

\bibitem{Georgi:1977tv}
  H.~Georgi and H.~D.~Politzer,
  \emph{Phys.\ Rev.\ Lett.}\  {\bf 40} (1978) 3;\\
%
  A.~M\'endez, A.~Raychaudhuri and V.~J.~Stenger,
  Nucl.\ Phys.\  B {\bf 148} (1979) 499;\\
%
  A.~K\"onig and P.~Kroll,
  \emph{Z.\ Phys.\  C} {\bf 16} (1982) 89;\\
%
  J.~G.~Chay, S.~D.~Ellis and W.~J.~Stirling,
  \emph{Phys.\ Rev.\  D} {\bf 45} (1992) 46;\\
%
  P.~M.~Nadolsky, D.~R.~Stump and C.~P.~Yuan,
  \emph{Phys.\ Lett.\  B} {\bf 515} (2001) 175
  [{\tt hep-ph/0012262}].

\bibitem{Oganesian:1997jq}
  K.~A.~Oganesian, H.~R.~Avakian, N.~Bianchi and P.~Di Nezza,
  \emph{Eur.\ Phys.\ J.\  C} {\bf 5} (1998) 681
  [{\tt hep-ph/9709342}];\\
%
  M.~Anselmino, M.~Boglione, A.~Prokudin and C.~T\"urk,
  \emph{Eur.\ Phys.\ J.\  A} {\bf 31} (2007) 373
  [{\tt hep-ph/0606286}];\\
%
  U.~D'Alesio and F.~Murgia,
  \emph{Prog.\ Part.\ Nucl.\ Phys.}\ {\bf 61} (2008) 394
  [{\tt arXiv:0712.4328}]. 

\bibitem{Collins:1981uk}
  J.~C.~Collins and D.~E.~Soper,
  \emph{Nucl.\ Phys.\  B} {\bf 193} (1981) 381,
  Erratum \emph{ibid.\  B} {\bf 213} (1983) 545.

\bibitem{Ji:2004wu}
  X.~D.~Ji, J.~P.~Ma and F.~Yuan,
  \emph{Phys.\ Rev.\  D} {\bf 71} (2005) 034005
  [{\tt hep-ph/0404183}];\\
%
  J.~C.~Collins, T.~C.~Rogers and A.~M.~Stasto,
  \emph{Phys.\ Rev.\  D} {\bf 77} (2008) 085009
  [{\tt arXiv:0708.2833}]. 

\bibitem{Collins:2008ht}
  J.~Collins,
  these proceedings
  [{\tt arXiv:0808.2665}]. 

\bibitem{Mulders:1995dh}
  P.~J.~Mulders and R.~D.~Tangerman,
  \emph{Nucl.\ Phys.\  B} {\bf 461} (1996) 197,
  Erratum \emph{ibid.\  B} {\bf 484} (1997) 538
  [{\tt hep-ph/9510301}];\\
%
  A.~Bacchetta, M.~Diehl, K.~Goeke, A.~Metz, P.~J.~Mulders and M.~Schlegel, 
  \emph{JHEP} {\bf 0702} (2007) 093
  [{\tt hep-ph/0611265}].

\bibitem{Boer:2003cm}
  D.~Boer, P.~J.~Mulders and F.~Pijlman,
  \emph{Nucl.\ Phys.\  B} {\bf 667} (2003) 201
  [{\tt hep-ph/0303034}].

\bibitem{Collins:1984kg}
  J.~C.~Collins, D.~E.~Soper and G.~Sterman,
  \emph{Nucl.\ Phys.\  B} {\bf 250} (1985) 199.

\bibitem{Ji:2006br}
  X.~Ji, J.~W.~Qiu, W.~Vogelsang and F.~Yuan,
  \emph{Phys.\ Lett.\  B} {\bf 638} (2006) 178
  [{\tt hep-ph/0604128}];\\
%
  Y.~Koike, W.~Vogelsang and F.~Yuan,
  \emph{Phys.\ Lett.\  B} {\bf 659} (2008) 878
  [{\tt arXiv:0711.0636}]. 

\end{thebibliography}
\end{document}